\documentclass[preprint,prc,superscriptaddress,unsortedaddress,showpacs,preprintnumbers,amsmath,amssymb]{revtex4}
\usepackage{tipa}

\usepackage{graphicx}
\usepackage{dcolumn}
\usepackage{bm}
\usepackage{color}

\def\beq{\begin{equation}}
\def\eeq{\end{equation}}
\def\bea{\begin{eqnarray}}
\def\eea{\end{eqnarray}}

\def\fun#1#2{\lower3.6pt\vbox{\baselineskip0pt\lineskip.9pt
  \ialign{$\mathsurround=0pt#1\hfil##\hfil$\crcr#2\crcr\sim\crcr}}}

\preprint{}

\begin{document}

\title{Neutron-proton pairing in Nuclear Matter}

\author{Xiao-Hua Fan}
 \affiliation{Institute
of Modern Physics, Chinese Academy of Sciences, Lanzhou 730000,
China}

\author{Xin-le Shang}\email[ ]{shangxinle@impcas.ac.cn}
 \affiliation{Institute
of Modern Physics, Chinese Academy of Sciences, Lanzhou 730000,
China}
\author{Jian-Min Dong}
 \affiliation{Institute
of Modern Physics, Chinese Academy of Sciences, Lanzhou 730000,
China}

\author{Wei Zuo}
\affiliation{Institute of Modern Physics, Chinese Academy of
Sciences, Lanzhou 730000, China}\affiliation{School of Nuclear
Science and Technology, University of Chinese Academy of Sciences,
Beijing 100049, China}

\begin{abstract}
The self-energy effect on the neutron-proton (np) pairing gap is
investigated up to the third order within the framework of the
extend Bruecker-Hartree-Fock (BHF) approach combined with the BCS
theory. The self-energy up to the second-order contribution turns
out to reduce strongly the effective energy gap, while the
\emph{renormalization} term enhances it significantly. In
addition, the effect of the three-body force on the np pairing gap
is shown to be negligible. To connect the present results with the
np pairing in finite nuclei, an effective density-dependent
zero-range pairing force is established with the parameters
calibrated to the microscopically calculated energy gap.
\end{abstract}

\maketitle

\section{Introduction}
About 60 years ago, the importance of pairing correlation in
nuclear systems were realized \cite{bohr}. Since then, a large
number of experimental data have been accumulated, supporting the
isovector (T=1) pairing between like nucleons \cite{fif,RMP1}.
However, no clear evidence supports the isoscalar (T=0) pairing
\cite{evi}, despite of the fact that the T=0 interaction is much
stronger than the T=1 interaction \cite{str}. The main suppression
of the T=0 pairing might result from the strong spin-orbit
splitting \cite{orbit1,orbit2}. And the recent calculations on the
Gamow-Teller transition also suggest that the T=0 pairing
interaction plays a decisive role for the concentration of
Gamow-Teller strength when the spin-orbit splitting becomes small
\cite{sun}.

On the other hand, the microscopic calculation of the T=0
neutron-proton (np) pairing with bare nucleon-nucleon interactions
in nuclear matter predicts a sizable energy gap with the magnitude
of 12 MeV \cite{bhf1,sh1,npd,sh2}, which seems too large to
reconcile with the empirical information available from finite
nuclei \cite{Fn2}. However, the microscopical predictions can be
significantly changed via a refinement of the theoretical framework,
{\color{red}such as considering the energy dependence of the
self-energy \cite{sig3,sig4}, including the relativistic effect
\cite{DBHF}, embodying the polarization effect \cite{sup1} and so
on}. In particular the {\color{red}polarization effect}
\cite{sup2,sup3,sup4,scr1,scr2} in nuclear medium has been shown to
enhance or quench the T=1 neutron-neutron (nn) pairing gap depending
on the nuclear environment. As for the T=0 np pairing, the recent
paper \cite{ulbd} indicates that the {\color{red}polarization
effect} exhibits much less significant effect for symmetric nuclear
matter at densities above the half of the saturation density. At low
densities, it remains difficult and an open problem. Another
significant re-scaling of the pairing gap in symmetric nuclear
matter may come from the dressing of nucleons in an interacting
system, which modifies the density of state and the effective energy
gap \cite{sig1,sig2}. These modifications result from the energy
denpendence of the single-particle (s.p.) self-energy
$\Sigma(k,\omega)$. The reduction of the gap due to the appearance
of a quasiparticle strength $Z$ factor is up to about fifty percent
for T=1 nn pairing in pure neutron matter \cite{sig1,sig2}, while it
may become as large as about seventy percent for T=0 np pairing in
symmetric nuclear matter \cite{sig3,sig4}.

In Ref. \cite{sig3}, within the framework of the Brueckner theory,
the effect of the energy-dependent self-energy $\Sigma(k,\omega)$
has been studied. However, the self-energy is calculated only to
the lowest-order approximation $M_{1}(k,\omega)$. As is known that
the imaginary part of $M_{1}(k,\omega)$ goes to zero below the
Fermi energy and the imaginary part of the \emph{rearrangement}
term $M_{2}(k,\omega)$ presents the contrary behavior (Im$M_{2}$
goes to zero above the Fermi energy) \cite{zuo0,zuo1,zuo2}. The
imaginary part of the self-energy also plays an important role in
predicting the energy gap \cite{sig3}. Therefore a more complete
investigation by including the effect of the rearrangement
contribution $M_{2}(k,\omega)$ is necessary in the study of the
T=0 np pairing within the extended BHF approach. In addition, the
three-body force (3BF) is expected to enhance the $^{3}PF_{2}$ nn
pairing at high densities \cite{dong}, but its effect on the T=0
np pairing has not been reported yet. A more accurate estimate of
the np pairing gap should include the 3BF effect.

In this work, we shall concentrate on the modification of the gap
equation including the energy dependence of the single-particle
self-energy $\Sigma(k,\omega)$ up to the third-order approximation
within the extended BHF approach. Moreover, the effect of 3BF on
the $^{3}SD_{1}$ np pairing is also considered. The paper is
organized as follows: In Sec. II, we briefly review the
self-energy within the extended BHF approach and the formalism of
the off-shell BCS gap equation. The numerical results and
discussion are shown in Sec. III, where an effective
density-dependent zero-range pairing force is provided with the
parameters fitted to the calculated energy gap. And finally a
summary is given in Sec. IV.



\section{Self-energy within the EBHF approach and the off-shell gap equation}
The present calculation of the self-energy $\Sigma(k,\omega)$ for
symmetric nuclear matter is based on the extended BHF approach,
for which one can refer to Ref. \cite{zuo1} for more details. The
microscopic 3BF supplement to the extended BHF scheme can be found
in Refs. \cite{tbf1,tbf2}. After several self-consistent
iterations, the effective interaction matrix G in the
Bruecker-Bethe-Goldstone (BBG) theory is obtained. Using the
G-matrix, the self-energy $\Sigma(k,\omega)$ can be calculated.


\subsection{Self-energy up to the third order}
\begin{figure}
\includegraphics[scale=0.8]{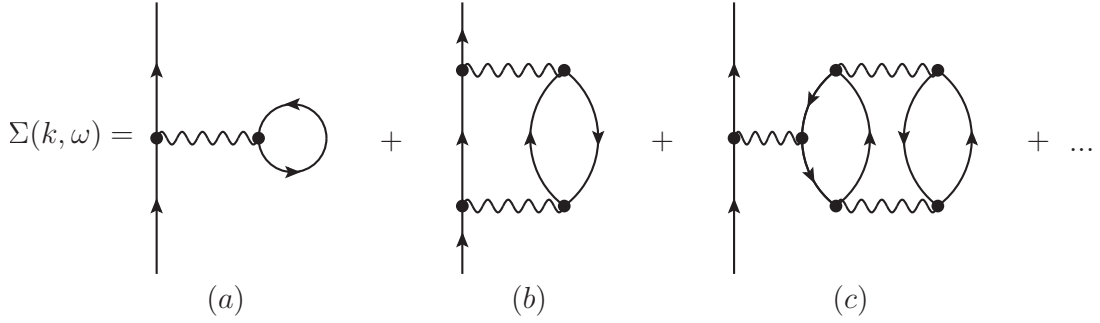} \caption{Hole-line expansion of the self-energy. }
\label{dig}
\end{figure}
Within the framework of the BBG theory, the self-energy
$\Sigma(k,\omega)$ can be expanded in a perturbation series
according to the number of hole lines \cite{ser}. The expansion up
to the third-order contribution is shown diagrammatically in
Fig.1. To the lowest-order approximation in the hole-line
expansion, i.e., the BHF approximation, the self-energy is written
as
\begin{eqnarray}
M_{1}(k,\omega)=\sum_{k'}n(k')\langle
kk'|G[\omega+\epsilon(k')]|kk'\rangle_{A},
\end{eqnarray}
where $\epsilon(k)$ is the s.p. energy spectrum in the BHF
approximation and $\omega$ is the starting energy. $n(k)$ is the
Fermi distribution function, which reduces to step function
$\theta(k-k_{F})$ at zero temperature. The subscript $A$ denotes
antisymmetrization of the matrix elements.

The next order in the hole-line expansion of the self-energy,
which is called \emph{rearrangement} term, is given by \cite{zuo0}
\begin{eqnarray}
M_{2}(k,\omega)=\frac{1}{2}\sum_{k'k_{1}k_{2}}[1-n(k')]n(k_{1})n(k_{2})\times\frac{|\langle
kk'|G[\epsilon(k_{1})+\epsilon(k_{2})]|k_{1}k_{2}\rangle_{A}|^{2}}{\omega+\epsilon(k')-\epsilon(k_{1})-\epsilon(k_{2})-i0}.
\end{eqnarray}
The corresponding diagrammatic sketch is shown in Fig.1 (b). It is
related to the particle-hole excitations in nuclear matter.

The third-order contribution in the expansion, as displayed in
Fig.1 (c), accounts for the fact that hole state $h'$ below the
Fermi surface is are partially unoccupied due to nucleon-nucleon
short-range correlations. Therefore, this contribution to the s.p.
spectrum is called the \emph{renormalization} contribution given
by \cite{zuo0,ren}
\begin{eqnarray}
M_{3}(k,\omega)=\sum_{h'}\kappa_{2}(h')\langle
kh'|G[\omega+\epsilon(k')]|kh'\rangle_{A},
\end{eqnarray}
with the lowest order of the depletion of the Fermi sea
\begin{eqnarray}
\kappa_{2}(h')=-[\frac{\partial}{\partial\omega}M_{1}(h',\omega)]|_{\omega=\epsilon(h')},
\end{eqnarray}
 where $h'$ refers to the hole state satisfy the condition
$|\overrightarrow{h'}|\leq k_{F}$. $\kappa_{2}(h')$ is the
probability that a hole state is empty. An estimate of
$\kappa_{2}(h')$ consists in using the average value of the
depletion, which is $\kappa=\kappa_{2}(h'=0.75k_{F})$
\cite{zuo1,ren}. Then the \emph{renormalization} contribution
$M_{3}(k,\omega)$ can be estimated by
$M_{3}(k,\omega)\approx\kappa M_{1}(k,\omega)$. Ref. \cite{ren}
also shows that $\kappa\sim 0.25$ for symmetric nuclear matter in
the density range $\rho\in(0.4\rho_{0},2.3\rho_{0})$, where
$\rho_{0}=0.17fm^{-3}$ is the empirical saturation density,
indicating the non-negligible effect of $M_{3}$.

\subsection{The off-shell gap equation and approximation}
Generally, the four-dimensional gap equation including the
self-energy $\Sigma(k,E)$ can be written as
\cite{gap1,gap2,gap3,gap4}
\begin{eqnarray}
\Delta(k,E)=\int\frac{d^{3}k'}{(2\pi)^{3}}\int\frac{d E'}{2\pi
i}\mathcal{V}(k,E;k',E')\Gamma(k',E')D(k',E'),
\end{eqnarray}
where the energy $E$ is defined as the energy relative to the
chemical potential $\mu$, i.e., $E=\omega-\mu$. And the kernel
$\Gamma$ is defined as
\begin{eqnarray}
\Gamma(k,E)&=&\mathcal {G}(k,-E)\mathcal {G}^{s}(k,E)\nonumber\\
&=&[\mathcal {G}^{-1}(k,E)\mathcal
{G}^{-1}(k,-E)+\Delta^{2}(k,E)]^{-1},
\end{eqnarray}
with
\begin{eqnarray}
\mathcal {G}(k,E)&=&[E-\frac{k^{2}}{2m}-\Sigma(k,\mu+E)+\mu+i0\ sign
E]^{-1},\\
\mathcal
{G}^{s}(k,E)&=&\frac{1}{{G}^{-1}(k,E)+\Delta^{2}(k,E){G}(k,-E)}.
\end{eqnarray}
The functions $\mathcal {G}(k,E)$ and $\mathcal {G}^{s}(k,E)$ are
the nucleon propagators in the normal state and in the superfluid
state for symmetric nuclear matter, respectively. We stress that
the neutron propagator differs only slightly from the proton
propagator due to the charge-dependent interaction and we ignore
this difference in this paper. The symmetry of $E$ in the kernel
$\Gamma$ is attributed to the time-reversal invariance of the
Cooper pairs.

In principle, the pairing interaction $\mathcal{V}(k,E;k',E')$
should include the {\color{red}polarization corrections}. In this
paper, the energy-independent interaction kernel is given merely by
the bare two-body potential or by the bare two-body interaction plus
a microscopic 3BF. Accordingly, the energy gap $\Delta$ is energy
independent as well. To be more precise, the angle-averaged gap
\cite{sh1,sh2,aag,aag2} equation in the $^{3}SD_{1}$ channel, which
actually has a couple channel structure involving a two-component
gap equation \cite{sd}, can be expressed as,
\begin{eqnarray}
\left(
\begin{array}{l}
\Delta_{0} \\
\Delta_{2}
\end{array}
\right)(k)=\frac{1}{\pi}\int_{0}^{\infty}k'^{2}dk'
 \left(\frac{1}{\pi}\int_{0}^{\infty} d E\
\textrm{Im} \Gamma(k',E)\right)\times\left(
\begin{array}{ll}
V^{00} & V^{02}\\
V^{20} & V^{22}
\end{array}
\right)(k,k')\left(
\begin{array}{l}
\Delta_{0} \\
\Delta_{2}
\end{array}\right)(k^{'}),
\end{eqnarray}
where $V_{LL'}(k,k')$ are the matrix elements of the bare
interaction in the relevant coupled channels $(L,L'=0,2)$ and the
total gap corresponds to
$\Delta^{2}(k)=\Delta_{0}^{2}(k)+\Delta_{0}^{2}(k)$.

 Resolving this gap equation exactly requires the knowledge of
the real and imaginary parts of the self-energy at arbitrary
energy $E$ and momentum $k$. A strong simplification may be
reached by assuming a small imaginary part of $\Sigma$,
$\textrm{Im}\Sigma\approx0$, which leads to a quasiparticle
approximation \cite{zuo2,zzf,shen}. In this approximation the
kernel $\Gamma(k,E)$ is an even function of energy: there exist
two symmetric poles $\pm\Omega_{k}$, corresponding to the
quasiparticle spectra in the superfluid state, on the real axis of
the complex $E$ plane. The integral over the energy range in the
Eq. (9) can be performed as follow:
\begin{eqnarray}
\frac{1}{\pi}\int_{0}^{\infty} d E\ \textrm{Im}
\Gamma(k,E)=-\frac{\mathcal {Z}_{k}^{2}}{2\Omega_{k}},
\end{eqnarray}
where the residue $\mathcal {Z}_{k}^{2}$ of the kernel at the pole
$\Omega_{k}$ corresponds to the quasiparticle strength which is
given by
\begin{eqnarray}
\mathcal {Z}_{k}^{-2}&=&\left(\frac{\partial\mathcal
{G}^{-1}(k,E)}{\partial
E}\right)\bigg|_{E=E_{k}}\times\left(\frac{\partial\mathcal
{G}^{-1}(k,-E)}{\partial
E}\right)\bigg|_{E=-E_{k}}\nonumber\\&=&\left[1-\frac{\partial\Sigma(k,\mu+E)}{\partial
E}\right]^{2}\bigg|_{E=E_{k}}.
\end{eqnarray}
The spectra of the single particle in normal state and the
quasiparticle in the BCS state are expressed as
\begin{eqnarray}
E_{k}&=&\frac{k^{2}}{2m}+\Sigma(k,\mu+E_{k})-\mu \nonumber\\
\Omega_{k}&=&\sqrt{E_{k}^{2}+\mathcal{Z}_{k}^{2}\Delta^{2}(k)}.
\end{eqnarray}
Then the gap equation can be approximated by
\begin{eqnarray}
\left(
\begin{array}{l}
\Delta_{0} \\
\Delta_{2}
\end{array}
\right)(k)=-\frac{1}{\pi}\int_{0}^{\infty}k'^{2}dk'\left(
\begin{array}{ll}
V^{00} & V^{02}\\
V^{20} & V^{22}
\end{array}
\right)(k,k')
\frac{\mathcal{Z}_{k'}^{2}}{2\sqrt{E_{k'}^{2}+\mathcal{Z}_{k'}^{2}\Delta^{2}(k')}}\left(
\begin{array}{l}
\Delta_{0} \\
\Delta_{2}
\end{array}\right)(k^{'}).
\end{eqnarray}
Note that the effective energy gap is $\mathcal{Z}_{k}\Delta(k)$
instead of $\Delta(k)$ due to the dispersive self-energy
\cite{zzf} which is also true for the exact gap of equation (9).
The presence of the quasiparticle strength, which is less than
unitary in a small region around the Fermi surface where the
Cooper pairs are mainly formed, reduces the pairing gap.

The gap equation should be solved self-consistently with the
density constraint since the pairing could modify the chemical
potential when the density is fixed in symmetric nuclear matter.
In the superfluid state, the density can be expressed as
\begin{eqnarray}
\rho=4\sum_{k}\int_{-\infty}^{0}\frac{d
E}{\pi}\textrm{Im}{G}^{s}(k,E),
\end{eqnarray}
where a factor of four comes from the spin and isospin degrees of
freedom.
In this paper the numerical investigation is based on a
self-consistently solution of the two coupled gap equations, Eqs.(9)
and (14). The self-energy is considered up to the third order in the
hole-line expansion.

\section{Results and discussions}
The numerical calculation here focuses on the $^{3}SD_{1}$ np
paring gap with inclusion of the self-energy effect and the 3BF
effect. The realistic Argonne $V18$  two-body interaction is
adopted as the pairing interaction which is consistent with the
self-energy calculated within the framework of the extended BHF
approach using the same interaction. And the microscopic 3BF
adopted here is constructed by using the meson-exchange current
approach as in Refs. \cite{tbf1,tbf2}.

\begin{figure}
\includegraphics[scale=0.8]{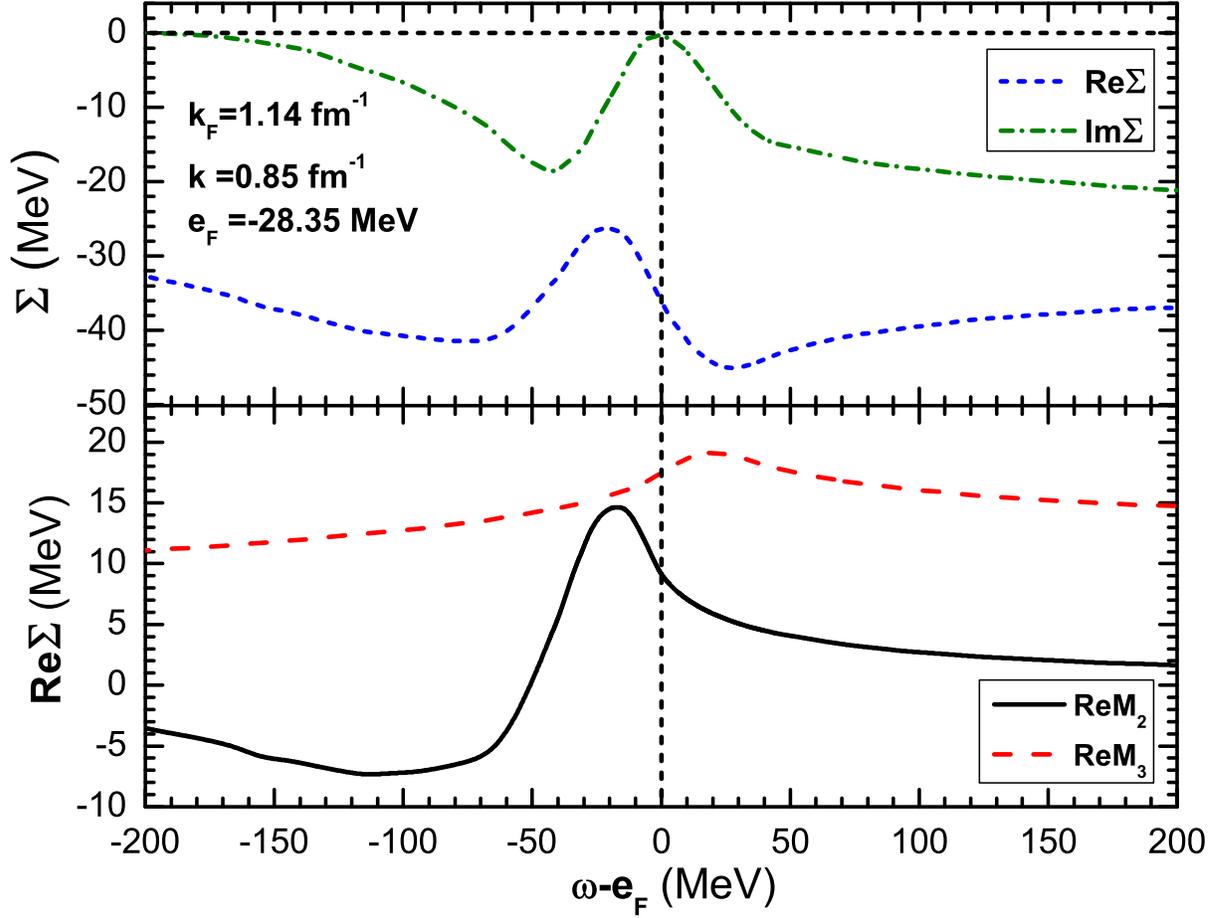} \caption{(Color online). The real and imaginary parts (upper panel)
of the self-energy $\Sigma$ and the the \emph{rearrangement} and
\emph{renormalization} terms (lower panel) of the self-energy
$\Sigma$ for $k=0.85fm^{-1}$ at $k_{F}=1.14fm^{-1}$. The
short-dashed vertical line indicates the position of the Fermi
energy.} \label{im3}
\end{figure}
As an illustrative example, the real and imaginary parts of the
off-shell self-energy $\Sigma(k,\omega)$ at a density of
$0.1fm^{-3}$ ($k_{F}=1.14fm^{-1}$) and a fixed momentum value of
$k=0.85fm^{-1}\approx0.75k_{F}$ is exhibited in the upper panel of
Fig.2. As mentioned in the introduction, the imaginary part of
$\Sigma(k,\omega)$ goes to zero at the Fermi energy. This is true
for the momentum $k=k_{F}$ as well, which implies the
quasiparticle strength approximation is reliable near $k_{F}$.
However, $\textrm{Im}\Sigma$ becomes sizable compared to the real
part of the self-energy at the s.p. energy when $k$ apart from
$k_{F}$, and the imaginary part should be handled seriously. The
lower panel of Fig.2 shows the real parts of $M_{2}$ and $M_{3}$.
The magnitude of $\textrm{Re}M_{3}$ is even larger than that of
$\textrm{Re}M_{2}$. Consequently, a reliable prediction of the
self-energy effect requires to account for the third-order
contribution $M_{3}$ and the second-order contribution $M_{2}$ at
the same footing.

\begin{figure}
\includegraphics[scale=0.8]{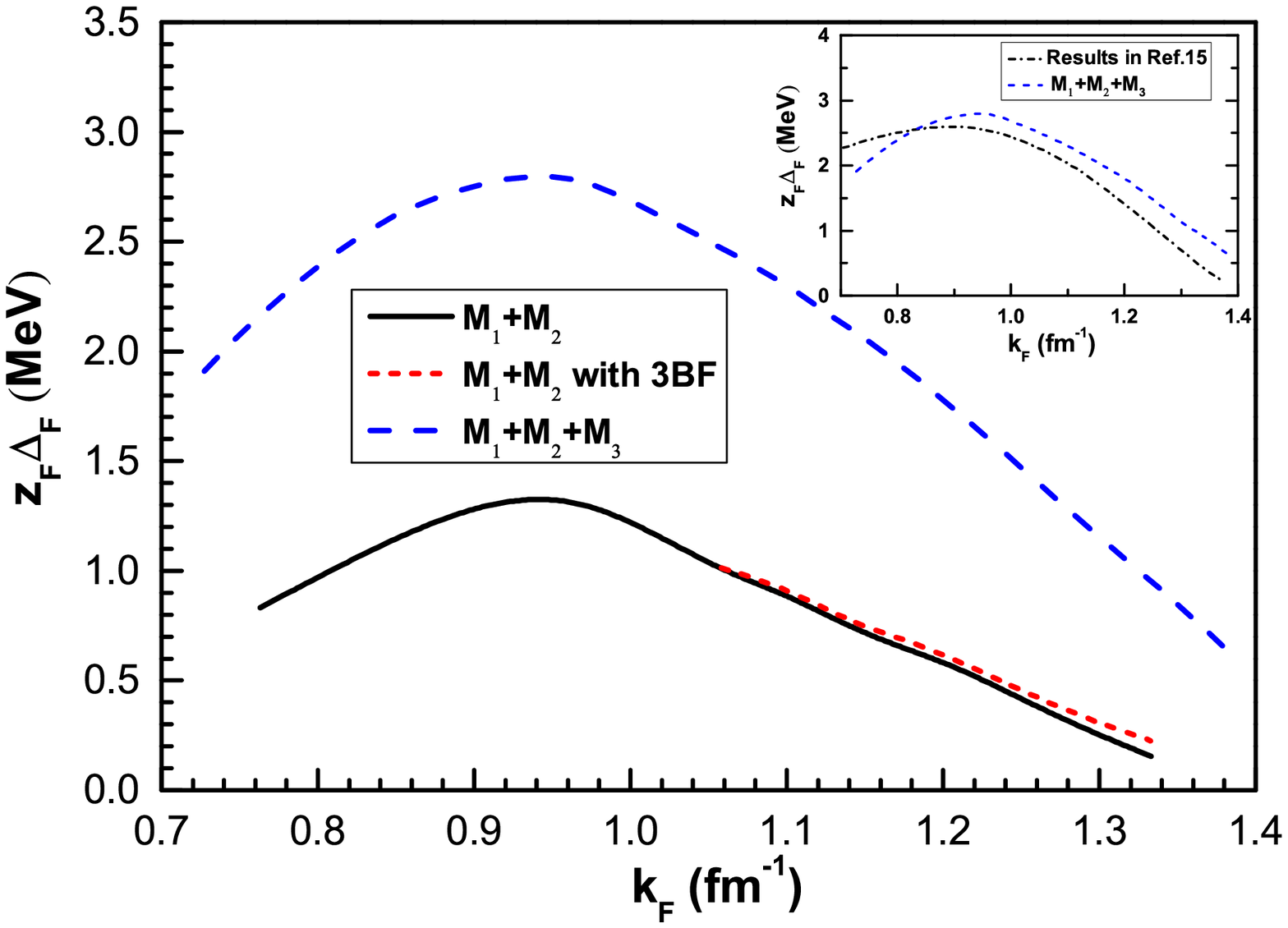} \caption{(Color online).
Neutron-proton effective energy gap in symmetric nuclear matter vs
the Fermi momentum $k_{F}$.
 The black solid, red short dashed and blue dashed lines
correspond to $M_{1}+M_{2}$ with 2BF, $M_{1}+M_{2}$ with 3BF and
$M_{1}+M_{2}+M_{3}$ with 2BF, respectively. The inset compares the
predictions in Ref. \cite{sig4} with the calculated result denoted
by $M_{1}+M_{2}+M_{3}$.} \label{gaps}
\end{figure}
The knowledge of $\Sigma(k,\omega)$ allows us to solve the gap
equations (9) and (14) exactly. Fig.3 displays the effective energy
gaps $\mathcal {Z}_{F}\Delta_{F}$ ($\Delta_{F}=\Delta(k_{F})$) at
the Fermi momentum $k_{F}$. In the calculated result denoted by
$M1+M2$ with 3BF, the 3BF contributions are embodied in both the
self-energy and the pairing interaction. It is shown that the 3BF
effect on the $^{3}SD_{1}$ np pairing is insignificant which is
consistent with the weak effect of 3BF on the equation of state of
nuclear matter at low density. The calculated results in the two
cases with and without $M_{3}$ by adopting the two-body force only
(the 3BF effect is negligible) are also compared in Fig.3. The
locations of the maximum $\mathcal {Z}_{F}\Delta_{F}$ with and
without $M_{3}$ are almost the same and around the density
$0.055fm^{-3}$ ($k_{F}=0.93fm^{-1}$). The behaviors of $\mathcal
{Z}_{F}\Delta_{F}$ with and without $M_{3}$ as functions of density
are nearly-identical as well. However, the magnitude of the
effective energy gap with $M_{3}$ is about two times of that without
$M_{3}$. The self-energy up to the second-order approximation leads
to a very strong reduction of the $^{3}SD_{1}$ np pairing gap, and
makes the pairing gap even smaller than the nn pairing gap in pure
neutron matter \cite{shen}. Nevertheless, the \emph{renormalization}
term of $\Sigma$ modifies both the quasiparticle strength and the
density of state, which finally enhances the paring gap
significantly.

{\color{red}In addition, a comparison between the present calculated
results within the extended BHF approach and the predictions within
the self-consistent in medium T matrix approximation \cite{sig4} is
shown in the inset. The present obtained effective gap in the
extended BHF approach turns out to be slightly larger than that in
the T matrix approximation in the density region of
$\rho>0.045fm^{-3}$, while it becomes a bit smaller than that in the
T matrix approximation when $\rho<0.045fm^{-3}$. In the extended BHF
approach, the maximal effective gap is located at $0.055fm^{-3}$
with a value of $2.8$MeV, and the effective gap $\mathcal
{Z}_{F}\Delta_{F}\approx 0.75MeV$ at the saturation density.
However, in the T matrix approximation, the maximal effective gap is
located at $0.05fm^{-3}$ with a value of $2.6$MeV, and the effective
gap $\mathcal {Z}_{F}\Delta_{F}\approx 0.45MeV$ at the saturation
density. Both calculations reveal a strong reduction of the
effective pairing gap at saturation density due to the effect of the
self-energy. The relativistic effect has been considered in Ref.
\cite{DBHF} and it may also lead to a strong suppression of the
pairing gap at saturation density.}

\begin{figure}
\includegraphics[scale=0.8]{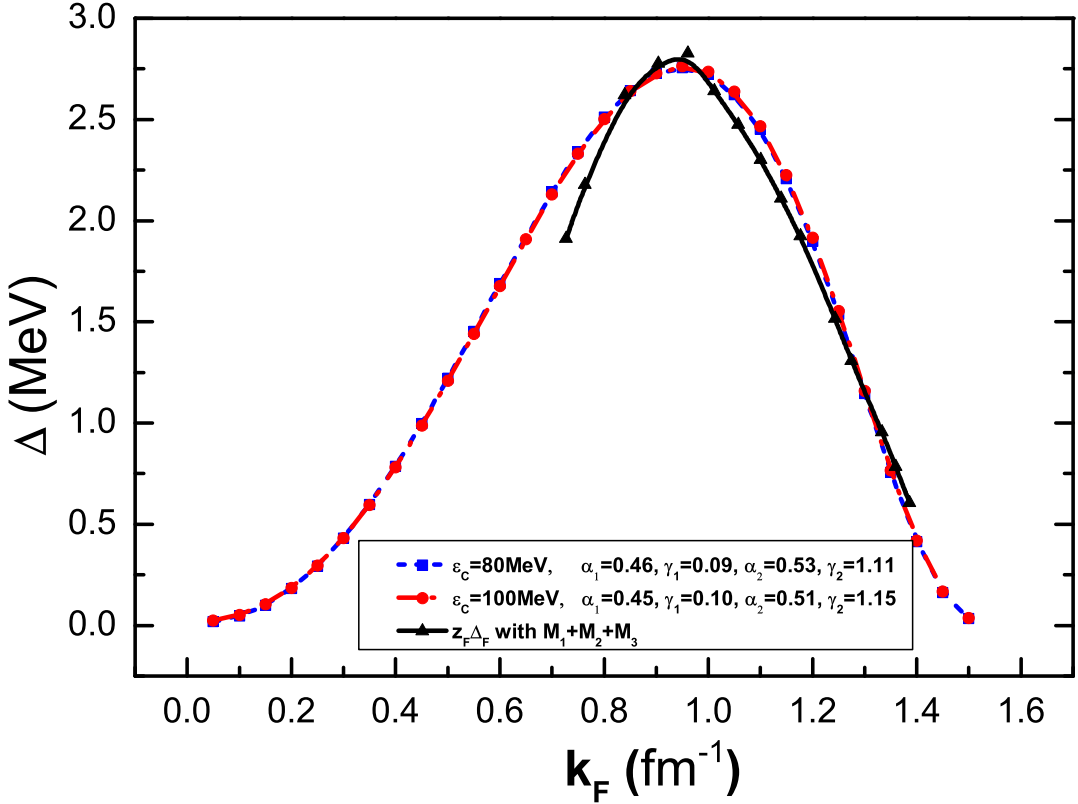} \caption{(Color online).
The np pairing gap using the effective density-dependent zero
range pairing force and the calculated effective energy gap data
within the EBHF approach.} \label{pair}
\end{figure}
To make contact with the pairing correlations in finite nuclei, we
propose an effective density-dependent zero-range pairing force
which include the reduction effect from the self-energy. The
parameters of the effective pairing force are determined to
reproduce the calculated gap values. We propose an effective pairing
force as follow:
\begin{eqnarray}
V_{pairing}(\textbf{r}_{1},\textbf{r}_{2})=v_{0}
\left[1-\alpha_{1}\bigg(\frac{\rho(\frac{\textbf{r}_{1}+\textbf{r}_{2}}{2})}{\rho_{0}}\bigg)^{\gamma_{1}}\right]
\left[1-\alpha_{2}\bigg(\frac{\rho(\frac{\textbf{r}_{1}+\textbf{r}_{2}}{2})}{\rho_{0}}\bigg)^{\gamma_{2}}\right]
\delta(\textbf{r}_{1}-\textbf{r}_{2}),
\end{eqnarray}
where $\rho_{0}=0.17fm^{-3}$ is the saturation density of symmetric
nuclear matter and
$v_{0},\alpha_{1},\gamma_{1},\alpha_{2},\gamma_{2}$ are parameters
which is adjusted to reproduce the present predicted gap in
symmetric nuclear matter. Being different from the proposed
effective density-dependent zero-range pairing force in Ref.
\cite{be}, we add an additional density-dependent factor
$\left[1-\alpha_{1}\bigg(\frac{\rho(\frac{\textbf{r}_{1}
+\textbf{r}_{2}}{2})}{\rho_{0}}\bigg)^{\gamma_{1}}\right]$ which is
expected to take into account the effect of $\Sigma$ on the pairing
gap. The exact physical picture of this additional term is not
clear, and it can improve the fitting which is exhibited in Fig.4
where the values of the obtained  parameters are also given. As is
well known, the zero-range pairing force must supplement a cutoff
$\varepsilon_{\textrm{c}}$. But in principle, the two parameters
$v_{0}$ and $\varepsilon_{\textrm{c}}$ are not independent and their
values should be chosen in such a way that the deuteron binding
energy is reproduced at zero density limit. Under this constraint,
the two parameters are determined to ensure the gap equations with a
solution of $\mu\rightarrow -1.12$MeV and a finite gap value when
$\rho\rightarrow0$. With this constraint, the values of the
parameter $v_{0}=-502.77$MeV and $v_{0}=-440.73$MeV correspond to
$\varepsilon_{\textrm{c}}=80$MeV and
$\varepsilon_{\textrm{c}}=100$MeV, respectively. The two groups of
parameters corresponding to the two different cutoffs produce nearly
the same gap values. {\color{red}It is worth noticing the shape of
the pairing gap as a function of $k_{F}$ (related to the density) in
Fig.4 looks similar to the behavior of the pairing gaps for
$^{1}S_{0}$ channel. In fact, except for the different dominant
density regions and magnitudes of the pairing gaps, the shapes of
the pairing gaps in $^{1}S_{0}$ and $^{3}SD_{1}$ channels behave
quite similar. As is known, the pairing correlation stems mainly
from the attraction of nuclear interaction at Fermi momentum
$k_{F}$. The attractions of both $^{1}S_{0}$ and $^{3}SD_{1}$
interactions at $k_{F}$ depends essential on density. And the
density behavior of both attractions are quite similar, i.e., first
increases up to a certain density and then decreases with the
density.}

\section{Summary and Outlook}
In conclusion, we have investigated the self-energy effect on the
$^{3}SD_{1}$ np pairing gap within the extended BHF approach plus
BCS theory. The \emph{rearrangement} and \emph{renormalization}
terms of self-energy are considered. The self-energy up to the
second-order approximation presents a strong reduction of the
effective energy gap, while the \emph{renormalization} term enhances
the pairing gap significantly. The maximum effective energy gap is
located at the density $\rho=0.055$ $\text{fm}^{-3}$ with a value of
$2.8$MeV. The 3BF effect on the np pairing gap is studied as well,
and it is found to be ignorable. Furthermore, an effective
density-dependent zero-range pairing force is proposed with the
parameters fitted to the calculated energy gap.

{\color{red}In this paper, we have concentrated on the self-energy
effect, and the polarization corrections to the pairing interaction
are not considered. Since an exact treatment of the polarization
effect is notoriously difficult due to its complicated, different
approximations are adopted to discuss the polarization effects
\cite{scr1,shen2}. It has been shown that the polarization effect is
negative to the pairing gap at low densities in the one-bubble
approximation, whereas it is slightly positive in the full RPA
limit. Moreover, as mentioned in the introduction, in Ref.
\cite{ulbd} it is indicated that the polarization corrections appear
to be negligible for moderate densities, yet it is still unclear and
an open problem at low density. An improvement of the present
calculation is to include the polarization effect in the future.}


\section*{Acknowledgments}
{The work is supported by National Natural Science Foundation of
China (No. 11435014, 11505241,11775276), the 973 Program of China
under No. 2013CB834405, the Youth Innovation Promotion Association
of Chinese Academy of Sciences.}
\appendix

\end{document}